# Of Decoherent Electrons and Disordered Conductors

P. Mohanty
*Condensed Matter Physics 114-36, California Institute of Technology, Pasadena, CA 91125*
*Department of Physics, Boston University, 590 Commonwealth Avenue, Boston, MA 02215*[1]

**Abstract**: Electron decoherence at low temperatures is important for the proper understanding of the metallic ground state, usually studied within the framework of Fermi liquid theory. It is also fundamental to various insulating transitions in low-dimensional disordered conductors, such as Anderson localization, which similar to the Fermi liquid theory relies on a vanishing decoherence rate at zero temperature. I review a series of interference experiments designed to study decoherence by a variety of intrinsic and extrinsic mechanisms. The goal is to determine if there is a truly intrinsic—and, in a sense, unavoidable—source of decoherence, coming from electron-electron interaction. In recent experiments we find that the saturation is not due to any mechanism involving magnetic impurities*. The interference experiments indicate that there is an ultimate source of decoherence even at zero temperature.

## 1. INTRODUCTION

Decoherence is the loss of coherence in a quantum system due to its coupling to an environment. Coherence of the quantum wave function is manifestly present in interference phenomena, which arise because of the linear superposition of wave functions corresponding to various alternatives in the evolution of the system [1]. In condensed matter physics, electrons inside the low-dimensional conductors behave quantum mechanically, very much like the photons or the electrons in a double-slit experiment [2-5]. The quantum mechanical behaviour appears because electrons in low-dimensional structures at low temperature can maintain phase coherence over long times, while traversing the entire length of the structure. This gives rise to additional interference corrections in the classical transport and thermodynamic properties of the conductor such as conductance and magnetization. The telltale signature of interference is the periodic oscillation of conductance [3] or magnetization [6] in the presence of a magnetic flux with a periodicity of the fundamental flux quantum h/e. Decoherence in these interference phenomena is depicted by the amplitude suppression of the periodic oscillations.

[1] Present address; Email: mohanty@physics.bu.edu.
* Recent experimental results from high-field measurements will be posted soon.

## 1.1 Electron decoherence

Electron decoherence arises because of the coupling of the electron to an environment capable of producing a time-dependent fluctuating electric field. Environments can either be intrinsic or extrinsic; intrinsic environments are phonons, magnetic, nonmagnetic and nuclear spins, and—most importantly—other electrons inside the conductor [7,8]. At millikelvin temperatures phonons freeze out, and decoherence due to the coupling of the electron to the phonon bath becomes negligible. It is possible to get rid of magnetic and nonmagnetic impurities down to a level of a part per million—that is, less than one impurity atom per a million host atoms. The electron decoherence rate is then not dominated by the magnetic or nuclear spins, and dynamical nonmagnetic impurities modelled as two-level systems. Scattering of the electron from these baths is further reduced at low temperatures due to the reduced occupation probability. The unavoidable mechanism of decoherence inside a clean conductor at low temperature is the interaction of the electron with other electrons. In a sense, all the material-dependent mechanisms can be labelled extrinsic. In the rest of the paper, we will consider electron interaction as the only intrinsic mechanism.

The above material-dependent mechanisms produce a fluctuating electric field that the electron couples to. In a similar fashion, the electron can also couple to an electric field originating outside the conductor. The foremost example of an extrinsic environment is the external high-frequency noise. Whether or not the dominating mechanism is extrinsic, it results in measurable effects in addition to the suppression of interference. An example of such effects is the heating of the electron due to nonlinear coupling, particularly in the presence of a strong electric field; since the electron temperature reveals this heating effect, discerning it from a genuine temperature-independent decoherence effect is crucial.

## 1.2 Decoherence and the physics of disordered systems

An important paradigm in condensed matter physics is the notion of quasi-particle. A number of phenomena can be understood in terms of weakly interacting quasi-particles though the Coulomb interaction between the electrons is rather strong. Interaction, in this framework, merely accounts for the renormalization of single-particle quantities such as the effective mass and the mean-field potential felt by the individual electrons. Screening of the electron interaction is affected if the size of the system is made smaller or if disorder is present. Nevertheless, the Fermi liquid framework of quasi-particles was shown to work in low dimensions and in the presence of disorder [9]. Recent experiments, however, have challenged this traditional picture. These include the topics of the current paper, saturation of decoherence time at low temperature [10], large persistent current in normal metals [11] and metallic states in two dimensions [12]; it appears that the single-particle picture may be inadequate in explaining these phenomena.

## 1.3 Role of decoherence in quantum information

The biggest challenge in quantum information science is the experimental realization of a quantum two-level system or a quantum bit (qubit) that can stay quantum mechanical for a time long enough for a number of manipulations [13]. The time over which the qubit remains quantum mechanical is known as its decoherence time. From a practical point of view, experimental investigation of decoherence is important to the success of quantum information science for the following two important reasons:
*(i) Measurement and manipulation time scale*: From a technical point of view, it is important to devise systems with long decoherence times so as to allow their study by conventional methods. In solid-state electronic systems, the time scale is known to be nanoseconds. That puts a limit on the speed (~ gigahertz) at which all manipulations have to be performed, even in these experimental prototypes of a single qubit.
*(ii) Scalability*: Even though it is possible to control and manipulate a single quantum two-level system, as has been recently shown, scaling to many qubits for entanglement and information processing appears to be a daunting task. The reason is that the effect of decoherence grows exponentially with the number of qubits in most models. With the current schemes, it will not be possible to go beyond a handful of qubits.

Understanding electron decoherence is the first step toward the realization of reliable quantum logic devices, and it necessarily precedes the steps of characterization, control, and reversal of decoherence of the qubits.

## 1.4 Organization of the paper

This paper discusses a series of interference experiments in mesoscopic systems. The goal of these experiments is to determine whether the temperature-independent decoherence rate is intrinsic or not. The notions of decoherence and interference in disordered conductors are briefly analysed in Section 2. Section 3 outlines various methods of determining decoherence rate, including weak localization, conductance fluctuations, Aharonov-Bohm effect and persistent current. Section 4 outlines the additional experiments that confirm that the observed saturation is indeed intrinsic, within the scope of the checks performed. These checks include the effect of magnetic impurities according to the standard Kondo effect, electron-electron interaction mediated by the magnetic impurities, electron heating, and high-frequency noise among others. Section 5 outlines the main points of the debate on whether quantum fluctuations of the electric field produced by electron-electron interaction can produce a finite decoherence rate.

# 2. INTERFERENCE AND DECOHERENCE

Two essential properties that distinguish a quantum mechanical system from a classical one are coherence and spin. Effects of spin are reflected in statistical properties such as noise and fluctuations whereas coherence is exemplified in interference phenomena. In condensed matter physics, interference effects are usually studied with the electron as the quantum system. These electrons remain quantum mechanical only for a finite time because of loss of coherence or decoherence induced by their coupling to the external world. Decoherence becomes less detrimental at low temperatures as the environmental degrees of freedom freeze out, or their thermal fluctuations, essential for decoherence, are suppressed. At low temperatures, in the millikelvin range, the typical time scale for electron coherence is of the order of nanoseconds. The distance the electrons traverse, while still remaining coherent, defines the phase coherence length. In experiments this length scale varies from a few microns down to a few nanometers. Thus the quantum-mechanical device—a solid-state device in which the electron behaves quantum mechanically—is usually nanometer scale in size.

## 2.1 Interference effects in mesoscopic conductors

The signature of quantum mechanics has always been interference—a direct consequence of the linear superposition principle. It postulates that if there are many alternatives for an event to occur, the net probability amplitude is the sum of the amplitudes for individual alternatives, and its square is the net probability. Consider, for instance, an event with two possible alternatives: an electron propagating through a solid with the geometry of a ring, shown schematically in Fig. 1(a). The electron could go through the upper branch or the lower branch. This set-up is actually the condensed-matter version of the two-slit experiment, the unavoidable picture that immediately comes to mind, the moment one speaks of interference. The quantum state of the electron $|\Psi\rangle$ is then the superposition of two states, the upper state $|u\rangle$ and the lower state $|l\rangle$ with complex amplitudes $\psi_{\text{upper}}$ and $\psi_{\text{lower}}$ respectively:

$$|\Psi\rangle = \psi_{\text{upper}}|u\rangle + \psi_{\text{lower}}|l\rangle \qquad (2.1)$$

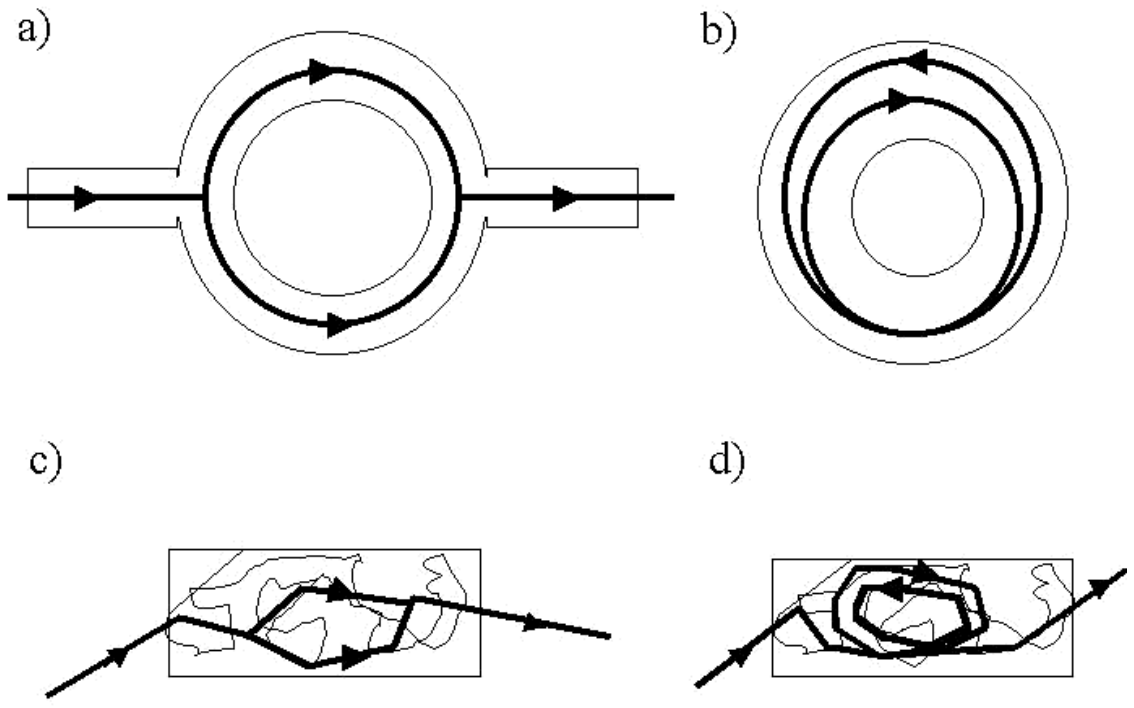

*Figure 1.* Four prominent manifestations of electron interference in the spirit of the two-slit experiment. Interfering pair of paths can involve either real paths (a and c) or time-reversed paths (b and d) corresponding to the characteristic flux scale of h/e or h/2e respectively. The schematics represent the following interference effects: (a) Aharonov-Bohm effect, (b) persistent current, (c) conductance fluctuations and (d) weak localization.

The probability of finding the electron at a point x on the right hand side of the ring is the magnitude squared of the projection of the electron state $|\Psi\rangle$ onto the position state $|x\rangle$:

$$|\langle x|\Psi\rangle|^2 = |\psi_{\text{upper}}|^2 |\langle x|u\rangle|^2 + |\psi_{\text{lower}}|^2 |\langle x|l\rangle|^2 + 2\,\text{Re}\left[\psi^*_{\text{upper}}\psi_{\text{lower}}\langle u|x\rangle\langle x|d\rangle\right] \qquad (2.2)$$

The last term represents interference between the two alternative ways for the quantum coherent electron to go from the left to the right. This term shows up in the conventional two-slit experiment as bright and dark fringes. In low-dimensional conductors, interference between various electronic paths is observed as quantum corrections to classical transport or thermodynamic properties such as conductance or magnetization.

Just as the fringe patterns exemplify constructive and destructive interference in the double-slit experiment, transmission of electrons inside a solid also displays interference, depending on the accumulated phase along their trajectories. An external magnetic field B changes the phase difference between two interfering paths by the amount

$$2\pi \frac{e}{h} \oint \vec{A} \cdot \vec{dl} = 2\pi \frac{e}{h} \oint B\vec{n} \cdot \vec{dS} = 2\pi \frac{\Phi}{h/e}, \tag{2.3}$$

where $\vec{A}$ is the vector potential. The acquired phase depends on the fundamental flux quantum h/e and on the flux enclosed by the paths Φ. Tuning of the so-called Aharonov-Bohm phase along the electron's path enables a variety of interference experiments inside a conductor. For instance, changing the phase from 0 through integer multiples of 2π allows the observation of a periodic oscillation in the electron transmission coefficient and is reflected in the conductance G of a disordered conductor with a ring structure:

$$G^{AB}(\Phi) = G^{AB}(\Phi + n\, h/e). \tag{2.4}$$

This is the Aharonov-Bohm correction. The interference of the complex set of paths inside the conductor gives rise to reproducible conductance fluctuations, and interference of time-reversed pairs of electron trajectories results in the weak localization of the electron in real space inside the conductor. Similar to various interference corrections to the conductance, thermodynamic quantities are also affected by interference, the prominent example being the persistent current in a normal-metal phase-coherent ring.

## 2.2 Electron decoherence due to environmental coupling

If the quantum system is coupled to an environment, then the total wave function of the system $|\Psi\rangle$ consists of both the system wave functions $|u\rangle$ and $|l\rangle$, and the environment wave function. Assume, for example, that the environment is in its ground state. The total wave function is

$$\left(\psi_{\text{upper}} |u\rangle + \psi_{\text{lower}} |l\rangle\right) \otimes |\chi_0\rangle. \tag{2.5}$$

This is a product state. After the interaction between the electron and the bath the total quantum state, however, goes into a state of "entanglement". The individual quantum states get entangled with the bath states which have changed after the interaction:

$$\psi_{\text{upper}} |u\rangle \otimes |\chi_1\rangle + \psi_{\text{lower}} |l\rangle \otimes |\chi_2\rangle. \tag{2.6}$$

The bath has now changed according to the states of the system, thus carrying information about the system. This leads to the loss of information or coherence (from the system to the bath), and hence decoherence. The probability density representing interference, the third term in the expression (2.2), now becomes

$$2\,\mathrm{Re}\left[\psi^*_{\mathrm{upper}}\psi_{\mathrm{lower}}\,\langle u|x\rangle\langle x|l\rangle\right] \rightarrow 2\,\mathrm{Re}\left[\psi^*_{\mathrm{upper}}\psi_{\mathrm{lower}}\,\langle u|x\rangle\langle x|l\rangle\langle \chi_1|\chi_2\rangle\right]. \qquad (2.7)$$

It is seen from this expression that the interference term will be completely suppressed if the bath states become orthogonal to each other after the interaction. Since the measurement of the interference only involves the system states, decoherence is quantized as the additional factor coming from the environment. This is the integration of the environmental degrees of freedom, which can be expressed in most cases as an exponential factor $e^{-t/\tau_\Phi}$. The rate at which the interference pattern is suppressed is the decoherence rate, and the time scale is the decoherence time $\tau_\Phi$. This also represents the time scale over which the two environmental states corresponding to the two system states become diagonal.

## 3. MEASUREMENT OF DECOHERENCE TIME

Quantitative estimate of decoherence involves the measurement of $\tau_\Phi$. However, the direct measurement of $\tau_\Phi$ as in a time-of-flight set-up is difficult, because it is of the order of nanoseconds. The second, important factor relates to the difficulty in measuring the electron probability density across a length, in analogy to the screen in the two-slit experiment: the fringe patterns cannot be directly measured in a conductor. In the two-slit experiment, the fringe pattern represents periodic oscillations of the probability density because of the varying path or phase difference along the screen. In mesoscopic conductors, on the other hand, the external modulation of the phase is achieved by a magnetic field. As discussed in Section 2.1, the Aharonov-Bohm flux modulates the phase difference between the upper and the lower paths, which meet at a single point. Any physical property expressed in terms of the electron transmission coefficient will then contain an interference correction, which will oscillate with the external flux. From the size of these oscillations, it is possible to quantitatively estimate the decoherence length or time. These fundamental Aharonov-Bohm oscillations are also contained in other effects such as conductance fluctuations, weak localization and persistent current—to be discussed in detail in this Section. A magnetic field corresponds to a magnetic time scale $\tau_H$ and length scale $L_H$. The decoherence time $\tau_\Phi$ is determined by comparing the interference effect with the pseudo-time $\tau_H$.

The translation of an interference effect such as weak localization into a time scale requires the understanding of how the time scale appears in the expression for this effect. In this Section, four different decoherence measurements are described. These are weak localization, conductance fluctuations, Aharonov-Bohm effect and persistent current. Note that two different measurements in the same sample may not necessarily yield the same value of $\tau_\Phi$ , as it is different in these different phenomena. In the next two subsections, we will list the time, length and energy scales in mesoscopic systems, and specific experimental aspects of magnetoresistance measurements for the determination of $\tau_\Phi$.

## 3.1 Mesoscopic dimensionality and time scales

The dimensionality of a mesoscopic system is very important as it determines the behaviour of interference effects. It is defined by the comparison of physical dimensions to the relevant length scales.

*Table 1.* Mesoscopic Length Scales

| Characteristic Length | Symbol |
|---|---|
| Average lattice spacing | $a_0$ |
| Electron mean free path | $l_e$ |
| Fermi wavelength | $\lambda_F$ |
| Thermal diffusion length | $L_T$ |
| Magnetic phase-breaking length | $L_H$ |
| Phase decoherence length | $L_\Phi$ |
| Sample dimensions (length, width, thickness) | L, w, t |

In a typical diffusive metallic conductor the average lattice spacing and the Fermi wavelength are on the order of a few angstroms. The electron mean free path $l_e$ depends on the degree of disorder, characterized by the diffusion constant $D = v_F l_e/d$, where $v_F$ is the Fermi velocity; the classical dimensionality d is defined with respect to $l_e$. Another way to determine D is to use the Einstein relation for the conductivity $\sigma = e^2 N(0) D$, where N(0) is the density of states at the Fermi energy, $E=E_F$. The magnetic length is defined as $L_H = \sqrt{3}\hbar / ewH$ , where w is the width and H is the magnetic field.

*Table 2.* Mesoscopic Dimensionality

| Dimension | Criteria |
|---|---|
| 0D | $L, w, t \ll L_H, L_\Phi$ |
| 1D | $\lambda_F \ll w, t \ll L_H, L_\Phi$ |
| 2D | $t \ll L_H, L_\Phi \ll w, L$ |
| 3D | $L_H, L_\Phi \ll t, w, L$ |

It is important to note that the one-dimensional systems defined are in fact “quasi-one” dimensional (in the sense that the notion of Luttinger liquids is not applicable here). In mesoscopic physics, the time scales are more fundamental than the length scales. In diffusive case, they are related by the diffusion formula, $L = \sqrt{D\tau}$ .

*Table 3.* Mesoscopic Time Scales

| Characteristic Time | Symbol | Relation to Corresponding Length Scale |
|---|---|---|
| Decoherence time | $\tau_\Phi$ | $\tau_\Phi = L_\Phi^2 / D$ |
| Magnetic time | $\tau_H$ | $\tau_H = L_H^2 / D$ |
| Thouless diffusion time | $\tau_D$ | $\tau_D = L^2 / D$ |
| Thermal diffusion time | $\tau_\beta$ | $\tau_\beta = L_T^2 / D = \hbar / k_B T$ |
| Mean free time | $\tau_e$ | $l_e/v_F$ |

## 3.2 Experimental aspects of decoherence measurement

The interference effects used for the extraction of $\tau_\Phi$ are extremely small, requiring an extremely high degree of care. In order to minimize or avoid various unwanted contributions, four important issues need to be properly addressed: (a) sample design, (b) sample fabrication, (c) measurement, and (d) analysis. These requirements are briefly detailed in the following.

a) For weak localization measurement, contributions from conductance fluctuations are avoided by making the length of the sample extremely long so that the sample contains many phase-coherent segments $L \gg L_\Phi$. The effect of voltage and current probes is manifested as an additional 2D contribution even in a four-probe configuration because of non-local effects. This is avoided by making the transverse dimensions of the voltage and current probes same as the quantum wire under study, and making the probes at least a few phase-coherent lengths long. A number of probes are desired to ensure the structural and material homogeneity of the sample. They also allow simultaneous weak localization and conductance fluctuation measurements. Furthermore, the size of the pads (normally designed for the ease of wire bonding) is important as they form as an antenna for high-frequency noise, causing both heating and decoherence. The separation distance between the pads should be such that they can capacitively short-circuit the high-frequency noise.
b) Proper sample fabrication of requires extreme care, as contamination of the sample by magnetic or nonmagnetic impurities is detrimental to the coherence effects. Starting material should be extremely pure, containing at best a few parts per million of other atoms. The level of magnetic impurity atoms should be below one part per million. Contaminants that are magnetic or superconducting should be avoided. Underlayers (normally used for the better adhesion of the quantum wire to the substrate) must be avoided. Contamination by exposure or during evaporation can also corrupt the quality of the sample.

c) Measurements of temperature dependence of $\tau_\Phi$ require an independent determination of the electron temperature. Possible heating and nonlinear effects by the measurement current must be avoided. Low-frequency switching and hysteresis effects can be checked for by the measurement of the noise spectrum.

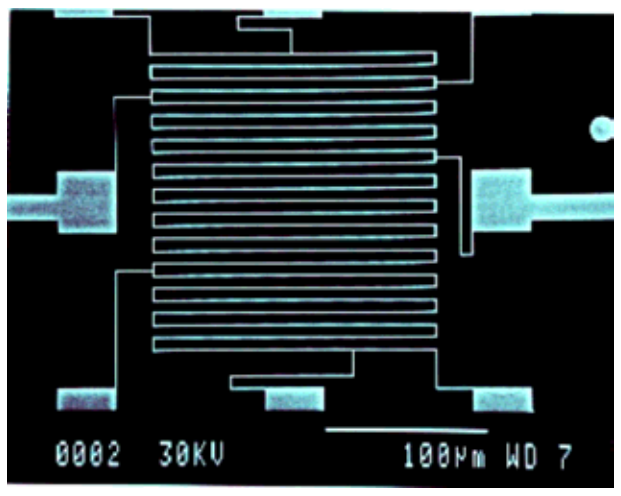

*Figure 2.* Scanning electron micrograph of a typical quasi-one dimensional wire.

d) A proper analysis is important for the determination of decoherence time and its temperature scaling. An accurate quantitative estimate of decoherence time requires a proper understanding of the interference effects, avoiding multi-parameter curve fitting to the data, and cross comparison with different kinds of interference measurements.

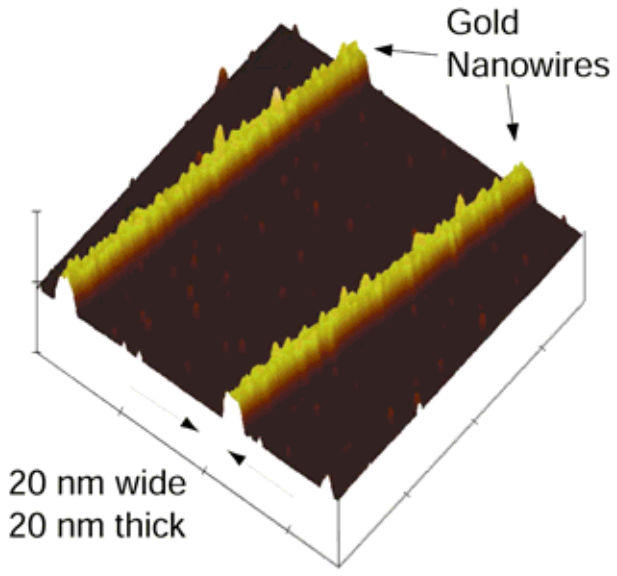

*Figure 3.* Atomic force micrograph of a section of a quasi-one dimensional gold wire.

## 3.3 $\tau_\Phi$ from weak localization

Weak localization arises due to the interference among time-reversed paths. Consider an electron inside a disordered conductor. It gets multiply scattered by the random medium from a momentum state **k** to a state **k'**, and another electron undergoes exactly the same sequence of scattering events as the first one. In real space, the electron propagates from one point **r** to another point **r'**. As shown in Figure 4(b), electron propagation from **r** to **r'** contains the trajectory of a self-intersecting path and another trajectory where the electron traverses the same path or another self-intersecting path close-by. In the absence of a magnetic field and spin-orbit scattering, the two equal contributions from time-reversed paths A add:

$$|A_1 + A_2|^2 = |A_1|^2 + |A_2|^2 + 2\,\mathrm{Re}(A_1^* A_2) = 4|A_1|^2 . \tag{3.1}$$

The return probability to the origin is $4|A|^2$, twice that of what is expected classically. The enhancement by quantum interference results in an enhanced probability of return for the electron, resulting in its localization. The application of an external magnetic field suppresses this interference correction. It is important to note that additional effects such as spin-orbit scattering can change the sign of localization. Thus weak anti-localization is observed in mesoscopic conductors made of gold in which spin-orbit scattering is strong.

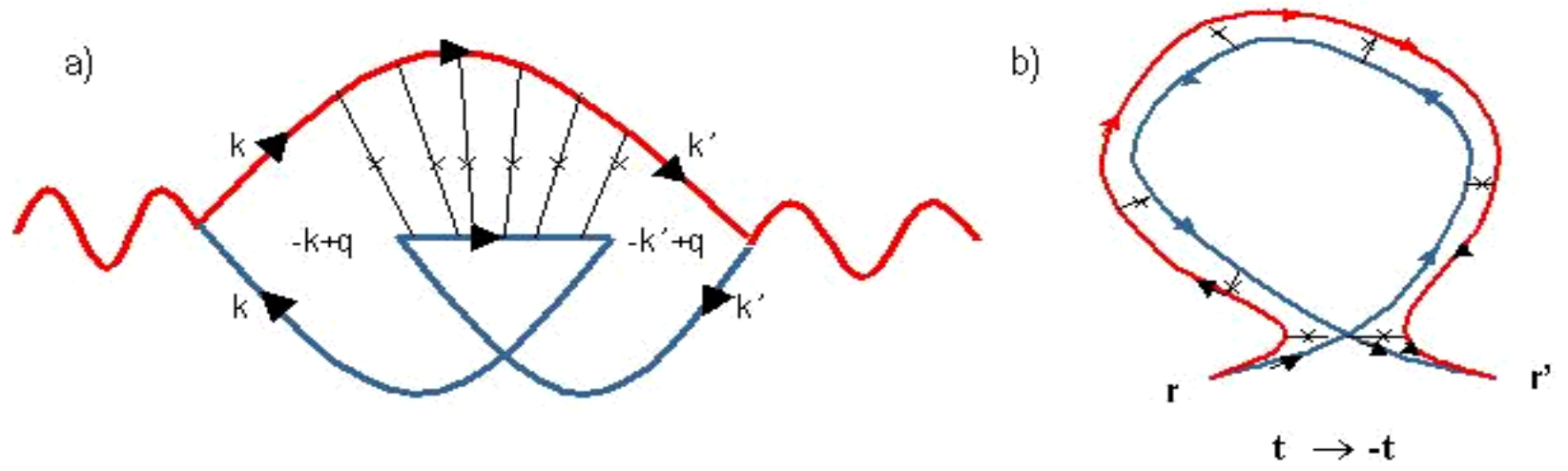

*Figure 4.* Weak localization diagram in (a) momentum space, and (b) real space.

Weak localization is quantitatively estimated by the probability that the electron returns to its original point. Here we follow Altshuler's description of weak localization [8]. The classical probability that the diffusing electron returns to a phase volume $dV$ at a time t is $dP = dV/(Dt)^{1/2}$. The phase volume is determined by the length of the tube $v_F dt$ formed by the electron's path and the cross-sectional area of the tube $\lambda_F^{d-1}$ (in d dimension). The return probability is related to the quantum conductance correction $\delta\sigma$:

$$\frac{\delta\sigma}{\sigma} = -\int dP = -v_F \lambda_F^{d-1} \int_\tau \frac{dt}{(Dt)^{d/2}} . \tag{3.2}$$

Application of a magnetic field introduces an Ahronov-Bohm phase, and the probability amplitudes are modified as

$$A_1 \to A_1 \exp(ie\overrightarrow{H}.\overrightarrow{S}/\hbar); A_2 \to A_2 \exp(-ie\overrightarrow{H}.\overrightarrow{S}/\hbar) . \tag{3.3}$$

The interference correction in the presence of a magnetic field H becomes

$$\frac{\delta\sigma(H)}{\sigma} \to -v_F \lambda_F^{d-1} \int_\tau \frac{dt}{(Dt)^{d/2}} \left\langle \cos(2\frac{e}{\hbar}\Phi) \right\rangle = \frac{e^2}{\hbar} \frac{(D\tau_\Phi)^{(2-d)/2}}{d-2} F(\frac{\tau_\Phi}{\tau_H}) . \quad (3.4)$$

The function F($x$)~ $x^{d/2-1}$ for $x$>>1 and it is a constant in the opposite limit. For d=2, the expression contains a logarithmic dependence. Thus $L_\phi$ or $\tau_\phi$ can be determined by fitting the magnetoresistance to the above-mentioned form with no other free parameter. ($\tau_H$ is fixed as the diffusion constant D and the width w are determined with a high degree of accuracy.)

### 3.3.1 Temperature dependence of $\tau_\phi$

Weak localization in a quasi-1D gold (Au) wire at 40 mK is displayed in Figure 5. Due to strong spin-orbit scattering in Au, weak anti-localization is observed instead of weak localization [4]. The size of the magnetoresistance dip at zero field is proportional to $L_\phi$. The typical correction to the resistance coming from weak localization is of the order of 0.1% in quasi-1D wires and 0.01% in 2D films. The data shown here is taken with the care and precautions listed in Section 3.2. The temperature dependence of $\tau_\Phi$ is extracted from a series of such weak localization traces measured at different temperatures.

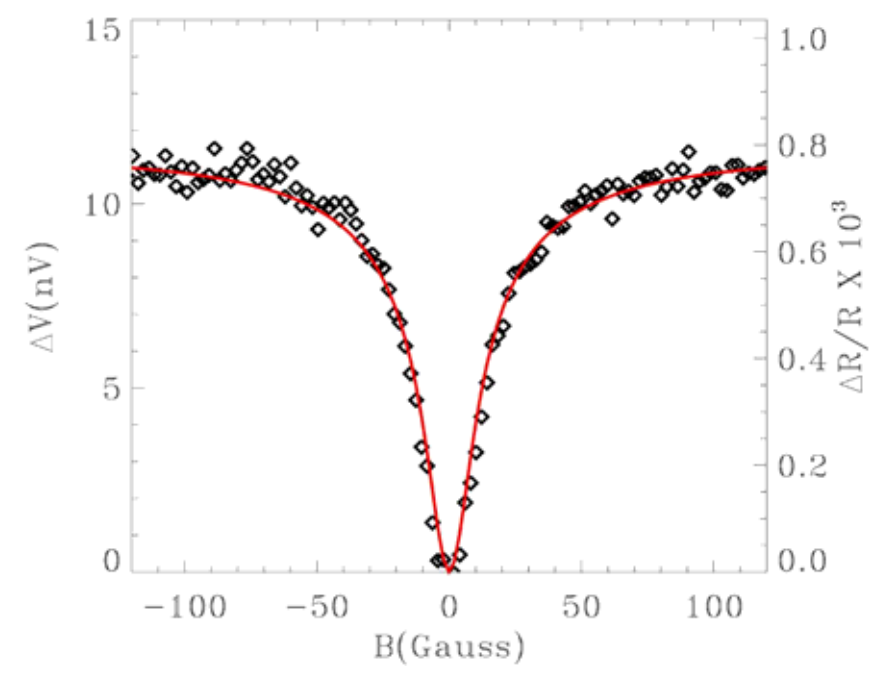

*Figure 5.* Magnetoresistance from weak antilocalization in a quasi-1D gold wire.

The temperature dependence of $\tau_\phi$ in quasi-1D gold (Au) wires for four representative samples is shown in Figure 6. At low temperatures all the samples invariably show the saturation of $\tau_\phi$. The temperature dependence below a temperature of 1 K deviates strongly from the $T^{-2/3}$ dependence expected from the conventional theory of electron-electron interaction. Furthermore, the saturation time $\tau_0$ and the temperature at which saturation onsets $T_0$ vary systematically with sample parameters as shown in Figure 6.

Low temperature saturation of $\tau_\phi$ has also been observed in a variety of 2D metallic films. In conventional theories $\tau_\phi$ is expected to vary as $T^{-1}$, which makes the contrast between the saturation and the expected temperature dependence rather strong. Both $\tau_0$ and $T_0$ can be tuned over orders of magnitude by changing the sample parameters. Figure 7 shows the temperature dependence of two films made from gold and gold-palladium with very different degrees of disorder. One shows the saturation at 4 K, the other merely shows a tendency towards the saturation as a deviation from $T^{-1}$ dependence [10].

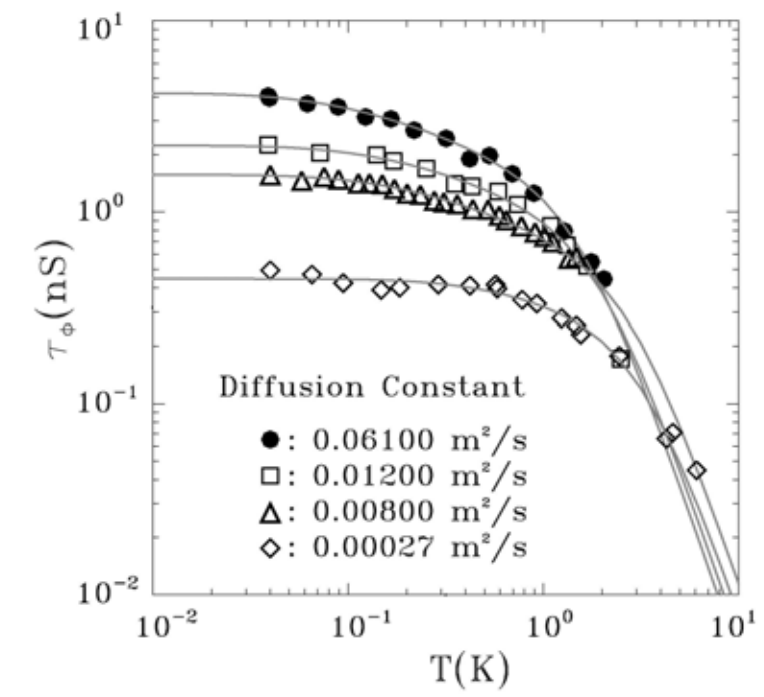

*Figure 6.* Saturation of decoherence time tF for four quasi-one-dimensional gold wires with varying degrees of disorder denoted by the diffusion constant D.

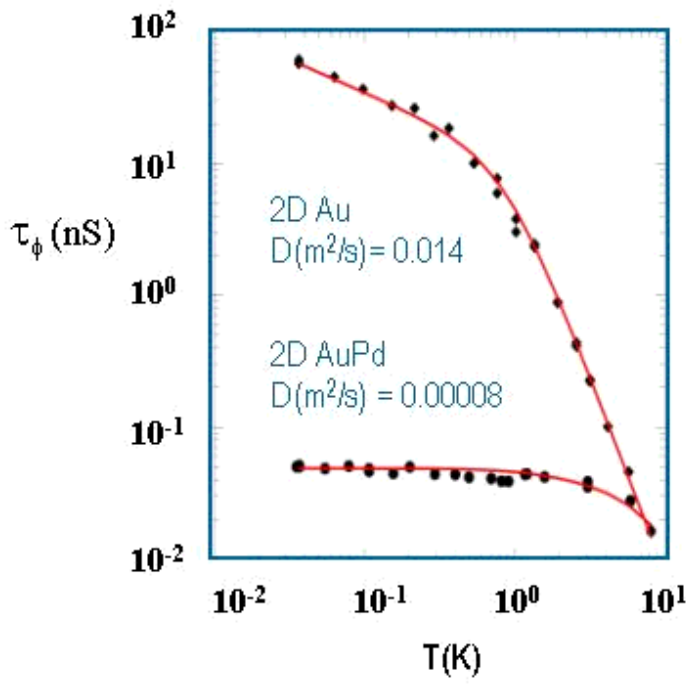

*Figure 7.* $\tau_\Phi$ in two-dimensional Au and AUPd films with very different degrees of disorder.

below 600 mK. By changing the diffusion constant D from 0.00008 $m^2/s$ to 0.0135 $m^2/s$, $\tau_\phi$ could be changed from 50 ps to ~ 60 ns in these 2D films and $T_0$ could be changed from ~ 1 K to below 20 mK.

Saturation of $\tau_\phi$ has been observed in many experiments on a wide range of mesoscopic systems [7]. These include quasi-1D and 2D films of Au [10], AuPd [14], Cu [15], and molecular AuPd wires [16] and semiconducting Si inversion layers [17], doped and undoped GaAs structures [18], 0-dimensional open GaAs quantum dots [19], and various 3D alloys [20,21]. Measurements on multi-walled carbon nanotubes [22] have displayed the saturation of the weak localization correction. Experiments on AuPd samples [16] in a wide range of widths also find the dependence of saturation on the sample parameters and the lack of a dominant contribution from a random mechanism, consistent with the experiments on Au wires. These experiments reinforce the earlier conclusion that the saturation is a real effect, most likely arising from electron-electron interaction [23].

In these experiments, the range of the saturation time $\tau_0$ extends over four decades [7], from few picoseconds to tens of nanoseconds. The temperature range of saturation in these experiments extends over three decades [7], from 20 K down to 20 mK. Controlled experiments show a clear trend between $\tau_0$ or $T_0$ and the sample parameters. In other words, with the appropriate choice of the sample parameters such as the resistance per unit length R/L, width, and diffusion constant, it is possible to tune $\tau_0$ and $T_0$. In a certain parameter range, $T_0$ can even be made lower than the lowest temperature of measurement (typically on the order of 10 mK or larger).

In spite of the strong evidence for the saturation of $\tau_\phi$, its ubiquity and universality, it is necessary to ensure that the effect is not due to artefacts. Recently, various extrinsic mechanisms contributing to the observed saturation have been proposed. The measurements described in this paper include extensive checks for various extrinsic mechanisms contributing to the saturation. We will discuss these experimental checks in Section 4.

## 3.4 $\tau_\Phi$ from conductance fluctuations

Reproducible conductance fluctuations arise from interference among the various complex paths inside the conductor [3,24]; the average of this correction does not vanish for a conductor of size comparable to or smaller than $L_\Phi$. Since all possible interference terms contribute to the conductance randomly, systems with identical parameters that characterize microscopic disorder, such as the diffusion constant, have very different conductances due to particular set of interference paths enforced by the particular impurity configurations. Measurement of conductance as a function of the external magnetic field allows the statistical mapping of the microscopic realization of the impurity configurations, because the interference patterns generated by the complicated set of paths is different for differing magnetic flux threading the paths. The characteristic correlation scale $B_c$ is given by the field required to thread a flux quantum through an area defined by $L_\Phi$ and the width w of the wire, $B_C \approx (h/e)/wL_\Phi$. In other words, from the correlation field scale it is possible to estimate $L_\Phi$:

$$L_\Phi^{CF} = C\frac{h/e}{wB_c}. \qquad (3.5)$$

The autocorrelation function of the conductance G (=1/R) is given by <G(B)G(B+ΔB)>, where the ensemble average is defined over a wide range of magnetic fields. The field scale over which this correlation function drops to half its maximum is $B_c$. The constant C varies from 0.95 for $L_\Phi >> L_T$ to 0.42 for $L_\Phi << L_T$; it is important to account for this crossover from the regime of dephasing-dominated smearing of fluctuations to the regime of smearing by energy averaging. A second method of extraction of the decoherence time $\tau_\Phi$ or the decoherence length $L_\Phi$ involves the determination of the rms value of the fluctuations over a large field scale. The standard theory of conductance fluctuations results in the following expression for the rms value:

$$\overline{\Delta G} \approx C_1\frac{e^2}{h}; \qquad (3.6)$$

the fluctuations are hence known as the "universal" conductance fluctuations. It assumes that the entire sample is phase coherent, $L<<L_\Phi$; for longer samples, the case in most experiments, there is an additional correction:

$$\overline{\Delta G} \approx C_2\frac{e^2}{h}\left(\frac{L_\Phi}{L}\right)^{3/2}; \qquad L_\Phi << L, L_T. \qquad (3.7)$$

The thermal diffusion length $L_T$ can be easily estimated. For a typical quasi-1D wire with diffusion constant of the order of 0.01 $m^2/s$, $L_T$ is on the order of a micron at 100 mK. In some cases, the necessary condition for the above asymptotic value, $L_\Phi << L_T$, may not be satisfied. A further correction arises because of thermal smearing if $L_\Phi$ exceeds $L_T$. At high temperatures, in the regime of $L_\Phi << L_T$, the total energy interval $k_BT$ is divided into intervals of width $E_C = \hbar D/L_\Phi^2$. This subdivides the system into N ~ $k_BT/E_c$ uncorrelated energy intervals, causing a further suppression of the fluctuations by averaging. The suppression of the fluctuations from their zero temperature or low temperature value is of the order of $1/\sqrt{N}$ ~ $L_T/L_\Phi$:

$$\overline{\Delta G} = C_3\frac{e^2}{h}\left(\frac{L_\Phi}{L}\right)^{3/2}\left(\frac{L_T}{L_\Phi}\right); \qquad \text{for} \quad L_T << L_\Phi << L; \qquad (3.8)$$

From detailed calculations the numerical values of the constants $C_1$, $C_2$ and $C_3$ are found to be 0.73, $\sqrt{12}$ and $\sqrt{8\pi/3}$ respectively.

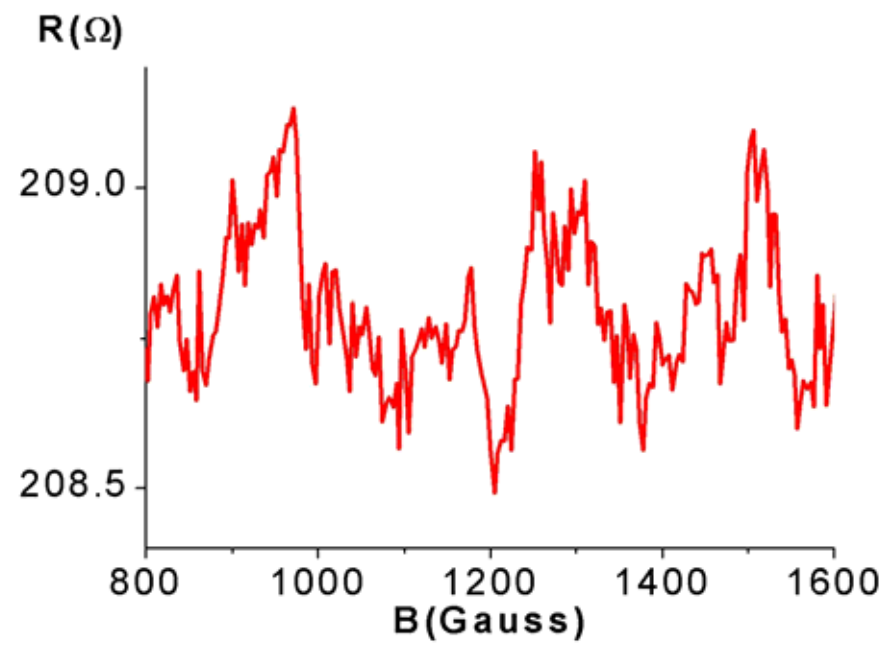

*Figure 8.* Reproducible conductance fluctuations in a quantum wire.

An explicit expression for $L_\Phi$ or $\tau_\Phi$ in terms of the measured fluctuation size $\overline{\Delta G}\big|_{EXPT}$ can be obtained:

$$\tau_\Phi^{CF} \equiv \frac{1}{D}\left( \overline{\Delta G}\Big|_{EXPT} \frac{h}{e^2} \frac{1}{C_3} \frac{L^{3/2}}{L_T} \right)^4 . \quad (3.9)$$

The temperature dependence of $\tau_\Phi$ extracted from conductance fluctuations is then measured by the rms value of the fluctuations. The advantage of this particular technique is that, in contrast to weak localization, $\tau_\Phi$ can be measured at high fields as well.

From a theoretical point of view, an important question arises with regard to the equivalence of $\tau_\Phi$ extracted from weak localization and conductance fluctuations. It has been recently shown that in the conventional framework the two time scales are the same [25].

The temperature dependence of $\tau_\Phi$ from conductance fluctuations can be determined by the rms value of the fluctuations or the critical field scale $B_c$ in the autocorrelation. The decoherence time measured by these methods shows the saturation at low temperature. It is an important result, because the measurements are taken at finite and large magnetic fields. Further discussions of conductance fluctuations are postponed to the section on the experimental checks.

## 3.5 $\tau_\Phi$ from the Aharonov-Bohm effect

A phase-coherent metallic ring threaded with a magnetic flux displays periodic Aharonov-Bohm (AB) oscillations with a periodicity of h/e, the flux quantum [2]. The periodicity is expected to be exact due to gauge invariance, $\Delta B = (h/e)/\text{Area}$, assuming no penetration of flux through the arms of the ring; the ring is phase-coherent since its size is smaller than the decoherence length $L_\Phi$. Phase-coherent contribution to the transmission of electrons the manifest as a periodic interference correction to the conductance: $G^{AB}(\Phi)=G^{AB}(\Phi+n\,h/e)$. The harmonics are denoted by the integer n, which represents the electron winding number around the ring, and Φ is the flux threaded through the area of the ring A, Φ= BA. The amplitude of the conductance oscillation due to this quantum-mechanical effect is on the order of the quantum conductance:

$$\Delta G^{AB} \approx K_1 \frac{e^2}{h}. \tag{3.10}$$

$K_1$ and the constants $K_n$ appearing in the rest of the sections are a constant of proportionality on the order of unity. For electrons with a finite decoherence time the size of the oscillations are reduced by an exponential factor even when the circumference of the ring L is smaller than $L_\Phi$. The amplitude of the AB oscillations for the nth harmonic is thus

$$\Delta G^{AB} \approx K_2 \frac{e^2}{h} e^{-\frac{nL}{L_\Phi}}. \tag{3.11}$$

At high temperatures, thermal averaging further reduces the amplitude, represented by the characteristic thermal diffusion length:

$$\Delta G^{AB} \approx K_3 \frac{e^2}{h}\left(\frac{L_T}{L}\right) e^{-\frac{nL}{L_\Phi}}. \tag{3.12}$$

If the spin-orbit scattering is weak, then $L_\Phi$ is modified according to $\tau_\Phi^{-1} \to \tau_\Phi^{-1} + 2\tau_{SO}^{-1}$ in the above formula, where $\tau_{SO}$ is the spin-orbit scattering time. In case of strong spin-orbit scattering, however, the amplitude suppression is exponential:

$$\Delta G^{AB} \approx K_4 \frac{e^2}{h}\left(\frac{L_T}{L}\right)\frac{1}{2}(3e^{-\frac{4}{3}\frac{L^2}{L_{SO}}} - 1)e^{-\frac{nL}{L_\Phi}}. \tag{3.13}$$

$L_\Phi$ or $\tau_\Phi$ can be determined from the amplitude of the AB oscillations using one of the above expressions in the appropriate regime.

Let us now consider a metallic ring of width w = 30 nm, thickness t = 20 nm and diameter = 1.7 μm. The expected periodicity of the h/e oscillation, the flux quantum divided by the area, is roughly 18 gauss. The AB oscillations with the expected periodicity in such a mesoscopic ring with resistance of 209 Ω at 40 mK are shown in Figure 9.

In the limit of high temperature and weak spin-orbit scattering one obtains an expression $\Delta G|_{EXPT}$,

$$\tau_\Phi^{AB} = -n^2 \tau_D \left[ \ln(\Delta G|_{EXPT} \frac{h}{e^2}\frac{L}{L_T} K_3 \right]^{-2}. \tag{3.14}$$

where $\tau_D = L^2/D$ is the electron diffusion time around the ring.

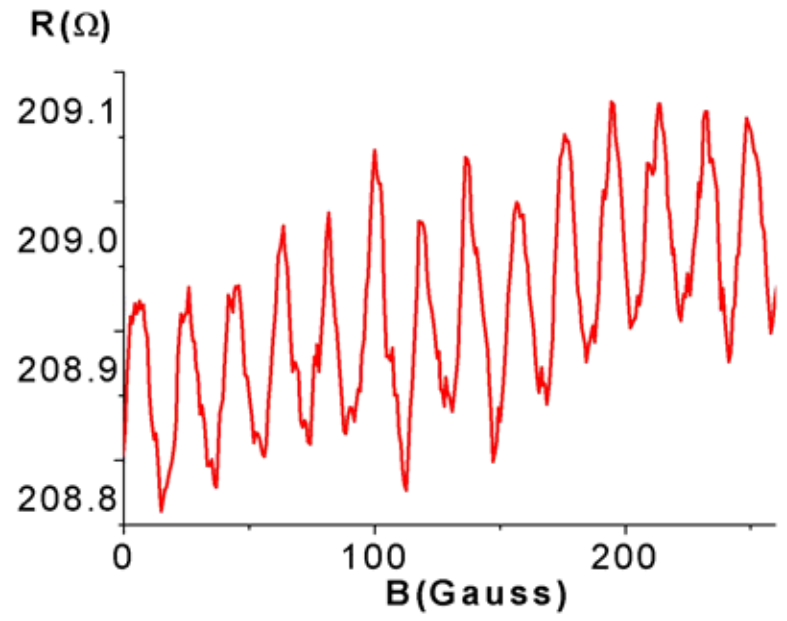

*Figure 9.* Aharonov-Bohm oscillations in a phase-coherent ring at 40 mK.

For the AB oscillations shown in Figure 9, one obtains the decoherence time $\tau_\Phi^{AB}$ by using the expression for the limit of weak spin-orbit scattering at low temperatures, the appropriate limit for this sample. However, the disadvantage of this method is the inaccuracy in the determination of $\Delta G|_{EXPT}$ from a finite number of oscillations. The decoherence time $\tau_\Phi^{AB}$ can also be obtained from the relative amplitudes of AB oscillations in higher harmonics as they decay according to $\exp(-nL/L_\Phi) \equiv \exp(-n\sqrt{\tau_D/\tau_\Phi})$. However, the study of the temperature dependence is not as accurate as it is with weak localization because (a) the AB oscillations in the data are not usually pure; (b) there are many crossover regimes depending on various length scales; and (c) it is hard to disentangle the contributions of conductance fluctuations and beating.

## 3.6 $\tau_\Phi$ from dissipative persistent current

Normal-metal phase-coherent rings exhibit persistent currents because of electron interference [5,6,11]. The presence of an Aharonov-Bohm flux Φ introduces a phase factor into the boundary condition for the electron wave function, $\Psi(x+L) = \Psi(x)e^{i2\pi\Phi/\Phi_0}$ such that all thermodynamic quantities are oscillatory in the applied flux with a period $\Phi_0$ = h/e. To minimize the free energy F, the isolated ring supports a persistent current, I(Φ) = -δF/δΦ, even in the presence of disorder. Furthermore, for an ensemble of rings, the fundamental harmonic of the current, periodic in h/e, is strongly suppressed due to its random sign in each ring. However, the harmonic, periodic in h/2e due to the contribution of time-reversed paths, survives both disorder and ensemble averaging. Figure 10 shows the h/2e component of the persistent current, measured in an array of 30 gold rings at 5.5 mK.

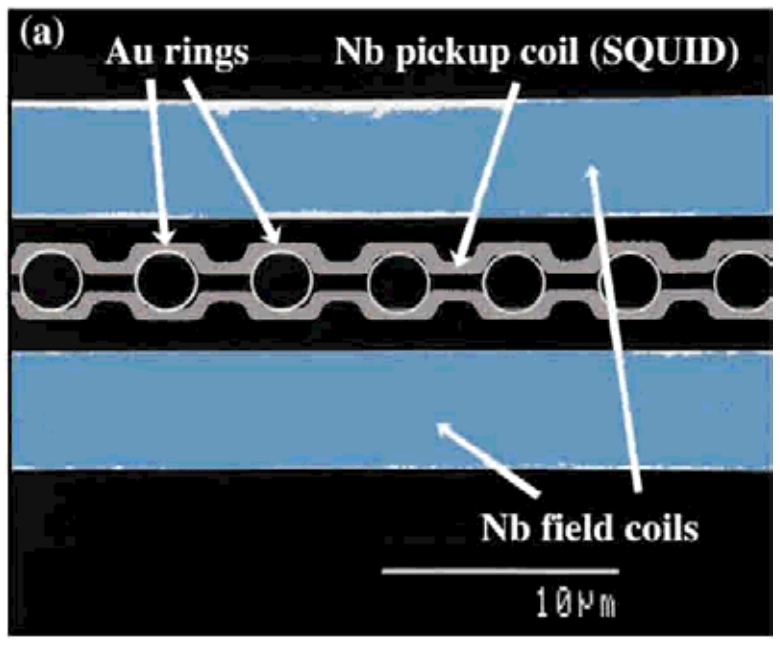

*Figure 10.* Scanning electron micrograph of a section of the array of phase-coherent rings.

Persistent currents can also result in an isolated phase-coherent ring of length ($L < L_\Phi$) due to the coupling of the electron to an intrinsic or extrinsic, high-frequency environment. In particular, an environment producing temperature-independent electric field fluctuations (in time) can result in a temperature-independent decoherence time $\tau_0$ and a large persistent current $I_{PC}$.

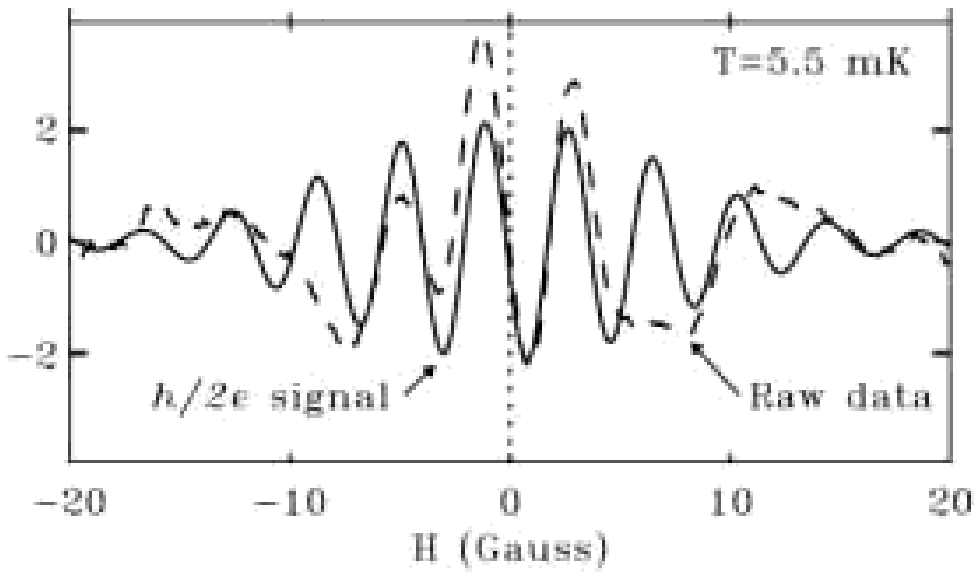

*Figure 11.* Persistent current in an isolated phase-coherent ring displays h/2e oscillations.

These two quantities are related to each other with a form specific to the environment [11,26]. For example, the current arising from a high-frequency noise source is given by

$$I_{PC}^{(n)} = C_\beta \frac{e}{\tau_\Phi} e^{-n\frac{L}{L_\Phi}}; \qquad (3.15)$$

n denotes the flux harmonic of the current, and $C_\beta$ is a constant of the order one. For a phase-coherent ring in the limit of $L \ll L_\Phi$, the magnitude of the current is inversely proportional to the decoherence time, and the saturation value is its maximum amplitude:

$$\tau_{\Phi(0)} \approx C_\beta \frac{e}{I_{PC}^{\max}}. \qquad (3.16)$$

In an experiment designed to study decoherence in persistent currents and weak localization, wires and isolated rings have been made with the same transverse dimensions (w and t) and fabrication conditions. The average persistent current per ring, measured in an array of 30 such gold rings, is found to be 0.06 nA for the h/2e component, corresponding to a decoherence time of $\tau_\Phi \sim 2$ ns. This is within a factor of 2 of the value of 4 ns obtained from weak localization measurements. The constant $C_\beta$ is roughly $2/\pi$.

A large persistent current is also generated from high-frequency fluctuations in a bath of two-level systems or even electron-electron interaction. Though these baths are intrinsic and at equilibrium, in contrast to the extrinsic nonequilibrium noise considered earlier, the Kravtsov relation [26] between the time and current is still valid. The temperature dependence of $\tau_\Phi$ can be determined from the persistent current measurements. The temperature decay is found to be exponential with the thermal length $L_T$ as the characteristic length, $e^{-L/L_T}$ ; decoherence time extracted from persistent current at low temperatures shows the saturation [11], consistent with the weak localization measurements on similar control samples [10].

## 4. EXPERIMENTAL CHECKS

In spite of strong evidence in favour of decoherence time saturation, it is necessary to ensure that the effect is not due to experimental artifacts, which may arise due to non-ideality in either the measurement or the sample. In the following we consider a series of extrinsic mechanisms, which cause an apparent saturation in the decoherence time at low temperatures. The goal is to discern these mechanisms from the intrinsic one, involving electron-electron interaction. Non-ideality in the sample may include magnetic impurity spins, nonmagnetic dynamical defects or two-level systems, coupling to nuclear spins etc. Measurement-induced artifacts include possible heating due to the measurement current or external high-frequency noise. Brief descriptions of these control experiments are given below. Taken in overall totality, they indicate that the observed saturation of decoherence time is most likely an intrinsic effect, coming from the unavoidable intrinsic electron-electron interactions.

### 4.1 Magnetic-impurity spins— the normal Kondo effect

Decoherence is enhanced due to the scattering of the electron by unwanted magnetic impurity ions such as iron, nickel, cobalt and manganese. The usual Kondo interaction between the electron and localized impurity spins has been ruled out by control experiments [10] in which an extremely small amount of iron ions (down to the level of 2 ppm) is introduced in gold wires after the measurement of $\tau_\Phi$ in pure samples.

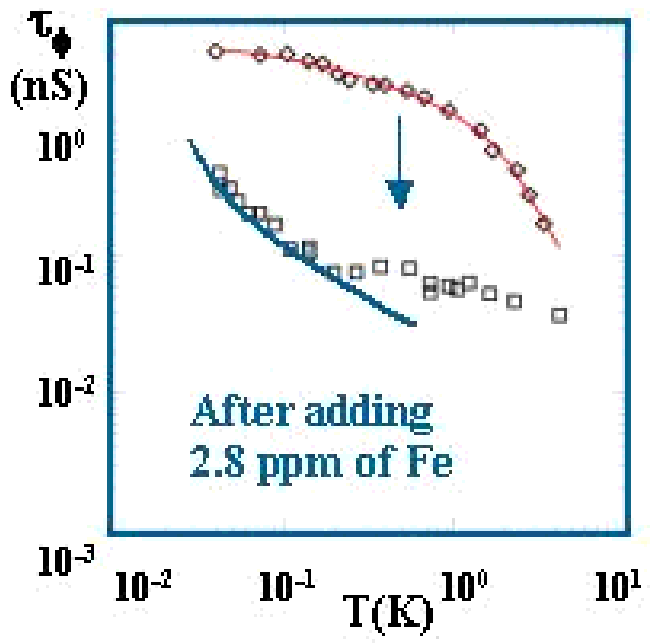

*Figure 12.* Effect of magnetic impurities on decoherence.

The magnetic scattering time $\tau_S$ has been extracted from the decoherence times with and without magnetic impurities. The excess scattering displays the anticipated peak at $T_K$. According to the normal Kondo physics, the strong temperature dependence $T^2$ of the scattering time below $T_K$ is due to the screening of the impurity spin by the electron cloud [27].

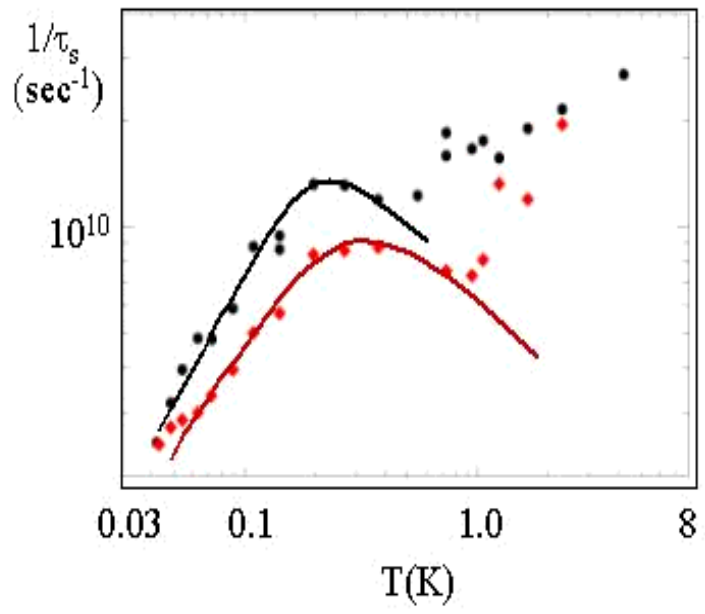

*Figure 13.* Magnetic scattering time in magnetically-doped wires.

The magnetic-impurity spins embedded in a host metal structure with mesoscopic dimensionality exhibit the normal Kondo physics. Though decoherence rate is enhanced and dominated by the additional magnetic scatterings, the temperature dependence is found to be very strong with the expected form $\sim T^2$. This rules out the possibility of strongly temperature-dependent Kondo scattering as the cause of the $\tau_\Phi$ saturation at low temperature.

## 4.2 Magnetic–impurity-induced electron interaction

Recently a novel saturation mechanism has been proposed, which invokes electron-electron interaction mediated by magnetic spins [28]. This mechanism requires an arbitrarily small number of impurity ions to generate an apparent saturation in decoherence at the level of nanoseconds. Even though the required concentration of impurity spins falls way below the spectroscopic level to be detected directly, there are many ways to test whether the observed saturation is indeed due to this particular mechanism.

This mechanism vanishes in the limit, $|\varepsilon = eV|$, $T \ll T_K$, where the electron energy $\varepsilon$ is given by either the temperature or the bias energy eV. Fermi liquid description holds in this limit, leading to the divergence of $\tau_\Phi$ as T goes to zero. However, for electron energies $\varepsilon \gg T_K$ the exchange interaction yields the following temperature-independent decoherence rate:

$$\frac{1}{\tau_0} = \frac{\pi}{2\hbar}\frac{n}{\nu} S(S+1)\left[\ln\left(\frac{eV}{k_B T_K}\right)\right]^{-4}, \tag{4.1}$$

where the concentration of impurity spins is given by n, the density of states at the Fermi level is denoted by $\nu$, and S is the spin of the impurity. The saturation time $\tau_0$ depends on the bias voltage V or current I.

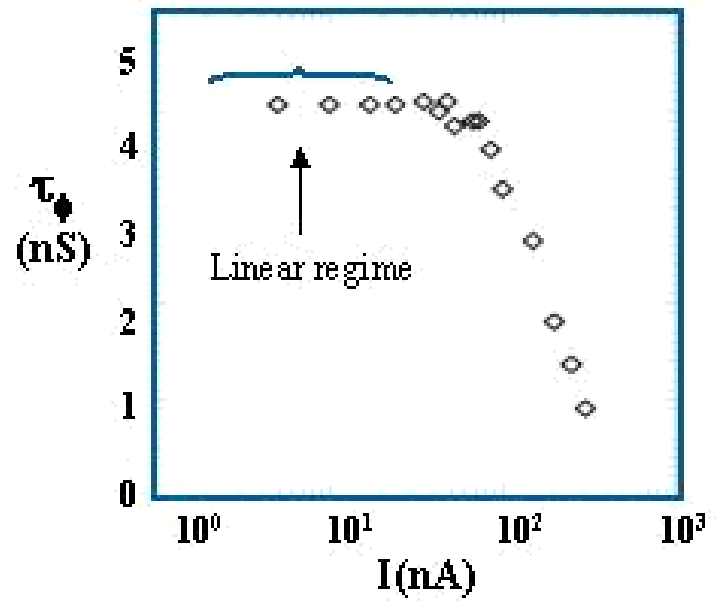

*Figure 14.* Dependence of $\tau_0$ on the bias current.

In the experiments, the decoherence time is usually measured in the regime in which it is independent of the bias current. Figure 14 displays the current dependence of the saturation time at 40 mK for a typical quasi-1D wire. In this particular sample $\tau_0$ is found to be independent of the bias over a decade up to 50 nA. Therefore the temperature dependence measured at a current, an order of magnitude smaller, is not governed by the above magnetic-impurity effect, or any nonlinear heating effect. The temperature dependence for all the samples shown in Figure 6 is taken in the low-bias regime. It is important to note that in all these samples $eV_\phi \ll T_K$, where $T_K$ is 0.3 K for iron impurities in gold. The saturation mechanism does not hold in this regime, though the measurement serves as a check for impurities with arbitrarily small $T_K$.

The magnetic field dependence of $\tau_\Phi$ serves as a check for mechanisms involving arbitrarily small number of magnetic impurities. At low fields, the electron energy is given by μB rather than eV as long as μB>eV. At high fields, the magnetic spins align in the direction of the field, and the random scattering process of decoherence is completely frozen. Figure 15 displays the temperature dependence of resistivity in an intentionally-doped sample. At a field of 2.5 tesla, a peak is observed around 2 K, indicating that at lower temperatures the iron spins are frozen out. At higher fields and lower temperatures, the decoherence time will not be dominated by magnetic scattering as confirmed in earlier experiments. Conductance-fluctuation measurements of $\tau_\Phi$ show saturation at low temperatures and high fields, strongly suggesting that the observed saturation in these samples is not due to mechanisms involving magnetic impurities[2].

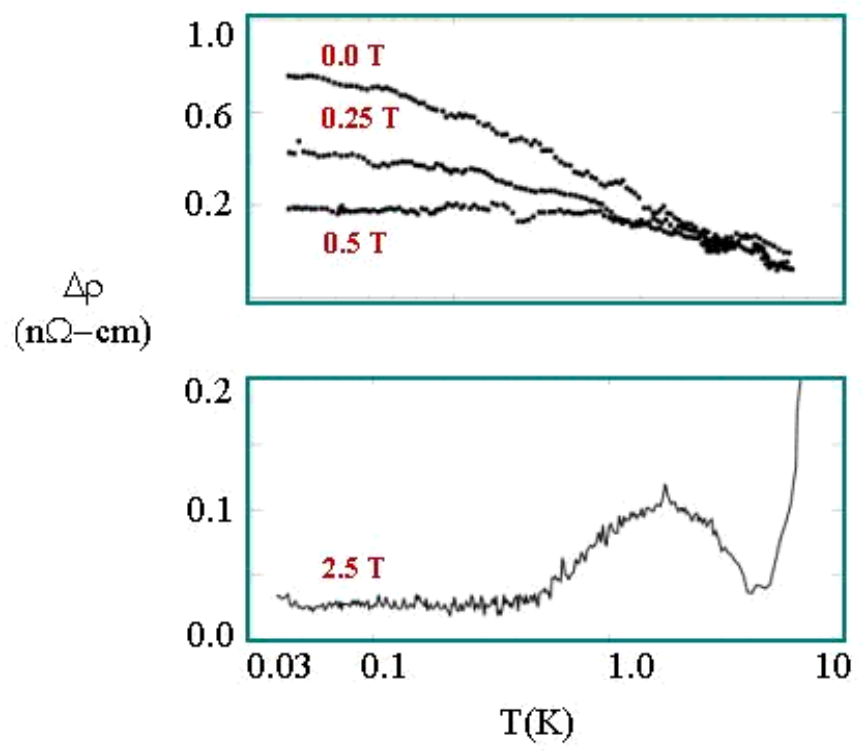

*Figure 15.* Freezing of magnetic spins at high fields in magnetically-doped samples.

Experiments on epitaxially-grown, high-purity GaAs heterostructures of quantum dots and quantum wires also show saturation of $\tau_\Phi$ at low temperatures. Recent measurement of weak localization in electron-doped high-$T_c$ superconductors also could not be due to magnetic impurities. Saturation in these experiments is not a random effect and it shows certain scaling with sample parameters, indicating that it is not from the presence of random magnetic impurities in the samples[3].

[2] *Post Script:* In addition to the fact that the Kaminski-Glazman mechanism is not applicable to the original experiments in which the condition $eV_\phi << k_BT_K$ was satisfied, recent conductance fluctuations measurements indicate that the observed saturation of $\tau_\Phi$ at high fields is not due to any mechanism involving any amount of magnetic impurity spins.

[3] Lin and coworkers have extensively probed the dominance of the mechanism in nominally pure samples by annealing measurements. They find an intrinsic source of saturation, and

### 4.2.1 Two-level atoms: 1/f noise

Extraneous decoherence is also caused by the electron coupling to dynamical non-magnetic defects, usually modeled by two-level systems (TLS). The resulting 1/f noise from the defects, dynamic on a time scale of nanoseconds, has been suggested as a possible saturation mechanism [29]. However, the required noise power level of $10^{-15}$ watts in the GHz range seems unlikely from the measurements done at low frequencies, as the expected noise power from an 1/f-noise distribution is on the order of a microwatt or larger below 1 Hz. Anticipated to be many orders of magnitude larger than the usual measurement power, 1/f noise power, switching or hysteresis from the TLS could not be detected down to a level of 1 nV per root hertz, suggesting their absence at the level required to give a saturation time in the range of nanoseconds. Dynamics on a slower timescale are expected to result in the scrambling of conductance fluctuations. This has not been observed in the experiments on the gold wires, suggesting that not only the required concentration of the TLS must be unreasonably high but also the power distribution has to be non-monotonic and highly unusual. In addition, the power level up to 1 GHz has to be considerably lower than a few femtowatts to be consistent with conductance fluctuations. Recent analyses of 1/f noise in metallic quantum wires and semiconducting quantum dots show that, to explain saturation in the nanosecond range, the required number or concentration of TLS has to be unreasonably high, a few orders of magnitude higher than what is found in metallic glasses[4].

### 4.2.2 Two-level atoms: the two-channel Kondo model

Another candidate mechanism involving TLS is the two-channel Kondo (2CK) model, where the interaction between the electron and the nonmagnetic spin (TLS) in the non-Fermi liquid regime gives rise to a temperature-independent scattering [30]. However, hysteresis and non-universality, anticipated from this mechanism, have not been observed in the experiments on the gold wires. Further analysis of the 2CK theory[5] finds a Kondo temperature much lower than the lowest temperature in the experiment for the 2CK model to be relevant to these experiments [31].

---

the mechanism is not the electron interaction mediated by an arbitrarily small number of magnetic impurities: Lin, Zhong and Li, Europhys. Lett. 57, 872 (2002), Lin and Bird (Review , to be published).

[4] Aleiner, Altshuler & Galperin, PRB 63, 201401 (2001); Ahn and Mohanty, PRB 63, 195301 (2001). Also see, Frasca, cond-mat/0112253; Schwab, Europhys. J. B 18, 189 (2000)

[5] Recent papers relevant to this issue are: Goeppert & Grabert, cond-mat/0105576 and PRB 64, 033301 (2001), Kroha & Zawadowski, cond-mat/0105026, and Goeppert, Galperin, Altshuler, and Grabert, cond-mat/0202353.

## 4.3 External high-frequency noise

The effect of ambient noise has been extensively studied by introducing into the cryostat calibrated high-frequency noise ($1/\tau_\Phi$ ~ GHz) and measuring its effect on both weak localization and electron temperature [32].

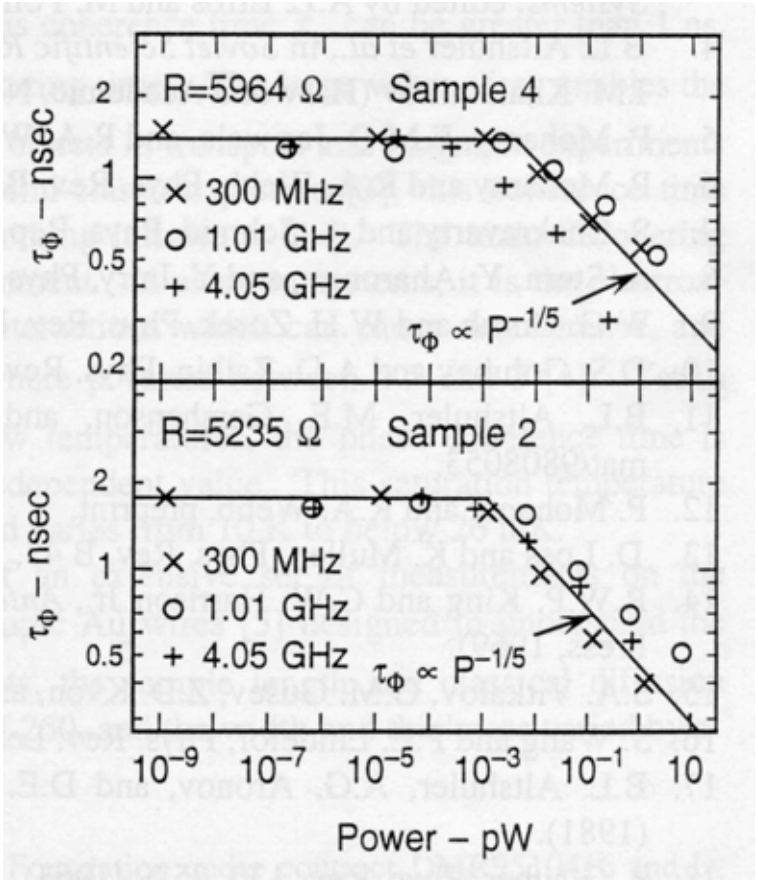

*Figure 16.* Effect of controlled high-frequency noise on decoherence.

The addition of external noise power at a single frequency results in extrinsic decoherence above a typical power level of femtowatts. However, much below this power level electron heating is observed in the resistance correction due to electron-electron interaction. This indication of electron heating prior to decoherence is possibly due to the fact that, at the particular frequencies, the coupling of the external bath to the electron is not optimal for decoherence. At higher powers decoherence is increased as a power law $\tau_\Phi \sim P^{-1/5}$ in agreement with theory.

As noise power is increased, substantial electron heating occurs prior to affecting decoherence time. Since no electron heating is observed in the absence of applied noise, ambient noise can safely be disregarded as a source of saturation in these experiments.

## 4.4 Electron temperature measurement

The low-frequency current applied for the transport measurement could result in a variety of anomalies, all of which have been experimentally checked [7,10]. Non-linearity (in the I-V curve) is avoided by restricting the excitation level to the low, linear regime. The applied bias across the phase-coherent length of the samples $L_\Phi$ is kept below the temperature, $eV_\Phi / k_B < T$, to ensure linear response. Hot electron effects due to non-equilibrium heating of the electrons above their equilibrium temperature are avoided; experimental checks to that end include the measurement of electron temperature with the electron-interaction correction—found to have the expected theoretical value and form ~ $T^{-1/2}$, by suppressing weak localization at a large, finite field [10].

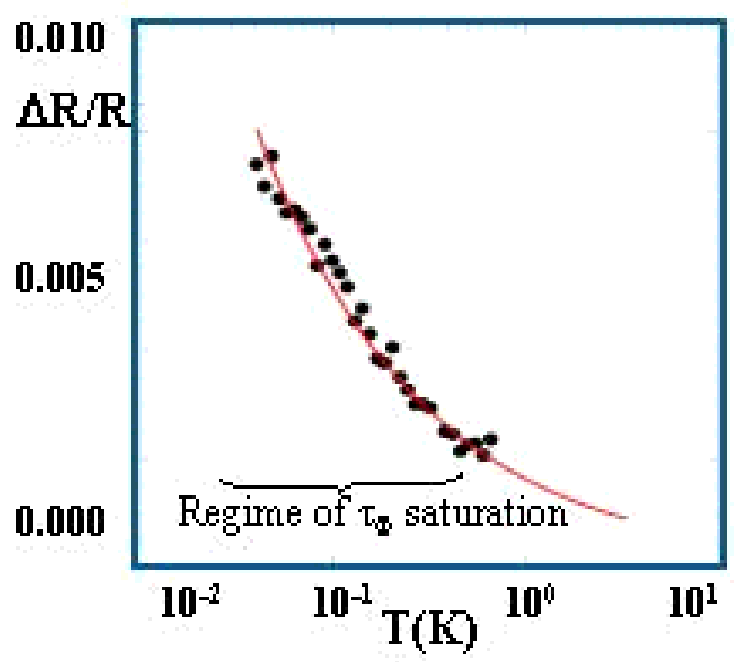

*Figure 17.* Electron temperature measurement using the resistance correction due to electron-electron interaction at a finite field. The curve indicates the expected theoretical form.

To further ensure the absence of electron heating due to the possible loss of thermal contact with the bath, calibration for the electron temperature is obtained additionally by the *a posteriori* destructive measurement of the Kondo effect—found to have the expected lnT dependence, and thermal Johnson noise measured by an integrated DC squid—found to have the expected noise power ~ $4k_BTR$ [33]. Most importantly, $\tau_\Phi$ is measured at the lowest temperature as a function of applied bias to verify that $\tau_\Phi$ is independent of the applied bias, as discussed earlier. This ensures the absence of additional detriments such as shot noise, and non-trivial configuration-dependent phonon effects. Anomalous contributions from weak localization in 2D contact pads, conductance fluctuations from the sample itself, and coupling of ambient high-frequency noise are minimized by the appropriate design of voltage and current probes as well as the sample length.

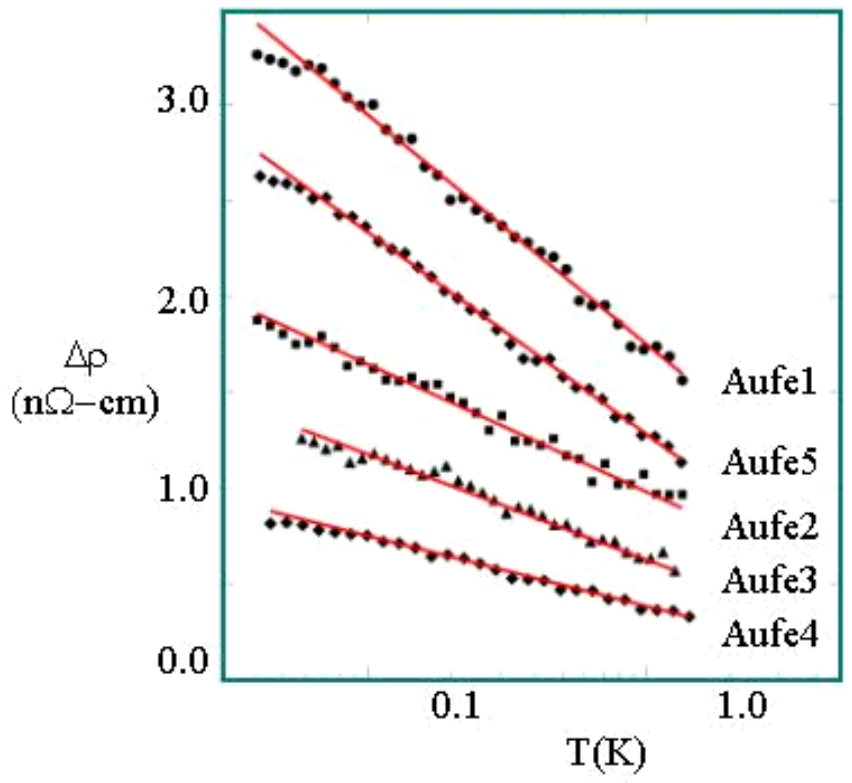

*Figure 18.* Electron temperature measurement by converting the samples into Kondo systems by magnetic doping. The straight lines indicate the expected lnT form.

## 4.5 Other extrinsic mechanisms

Interaction with nuclear magnetic moments, suggested as a possible mechanism, results in decoherence rates six to eight orders of magnitude smaller than the measured values. Furthermore, the apparent lack of material dependence argues against this mechanism as well as other material-dependent mechanisms such as the Stoner instability in 2D systems, not expected in gold. As discussed earlier, recent experiments on 3D samples made from various materials, and AuPd wires and films also reveal the material-independent but sample parameter-dependent saturation. For the unequivocal verification of the saturation in a wider range of thermal times $\hbar/k_B T$, more direct experiments are needed on a set of samples with a range of decoherence times.

# 5. ROLE OF ELECTRON INTERACTION AND QUANTUM FLUCTUATIONS[6]

The only intrinsic mechanism that can be common to any electronic system is the electron-electron interaction. This Coulomb interaction among the electrons can be represented as a fluctuating electromagnetic field with both transverse and longitudinal modes. It was initially suggested that the saturation of decoherence could be due to the quantum fluctuations of the intrinsic electromagnetic field [23]. In fact, the thermal fluctuations of this electric field give rise to the so-called classical Nyquist mechanism in the conventional theory for electron interaction at high temperatures. The essence of the proposal is decoherence by the quantum Nyquist mechanism involving quantum fluctuations of the electric field. This proposal is consistent with the density-matrix approach, usually implemented in a system-bath model, such as the Caldeira-Leggett model. However, both the idea [23] and the detailed theories [34] of decoherence by quantum fluctuations have been heavily debated both on concepts and the formal aspects of the calculations [35]. In order to present the two sides of the controversy, we outline the main conceptual arguments against possible decoherence and the counter arguments in the following[7].

[6] Since the submission of the article, a number of papers relevant to this section have appeared: Guinea (cond-mat/0112099), Belitz and Kirkpatrick (cond-mat/0111398 and 0112063) and comment by Golubev and Zaikin (cond-mat/0111527), Golubev, Zaikin and Schon (cond-mat/ 0110495) and Aleiner, Altshuler and Vavilov (cond-mat/0110145).

– An electron or any particle for that matter with energy $k_BT$ cannot excite the high-frequency quantum modes of the environment or the electric field. The role of these high frequency modes is to provide a temperature-dependent renormalization of the static disordered potential. Thus, for all practical purposes, the environment at T=0 remains inactive and induces no decoherence. The counter argument involves the example of a particle coupled to a bath of harmonic oscillators with a coupling linear in the coordinates of the bath oscillators. Coupling to the particle does not excite the individual bath oscillators when they are in their respective ground states at T=0. However, the linear coupling shifts the origin of the harmonic oscillators. This creates a back reaction on the particle causing both decoherence and dissipation, consistent with the results of an exactly solvable Caldeira-Leggett model of ohmic bath. Well known for almost two decades, calculations show that the off-diagonal elements representing coherence decay as a power law in time with the exponent set by interaction.
– Understanding of metals involves an ideal many-body ground state at T=0. Thus at T=0 there cannot be any scattering, as the many-body system resides in its ground state. The counter argument is the ill construction of the argument itself. The many-body system consists of a single particle (electron), whose coherence properties are measured in a transport experiment, and the rest of the electrons. Without this division of the whole system into a system of interest and the remainder or environment, even the notion of decoherence is meaningless. The whole many-body system is closed and quantum mechanical, whereby pure states transform into pure states unitarily. A pure state of the closed combination is compatible with each part being in mixed states. Decoherence is obtained by considering the density matrix operator of the combination, and tracing out the "irrelevant" part or the environmental degrees of freedom. Thus, one can indeed have a pure ground state for the many-body system at T=0 and a partially-decohered (finite $\tau_\Phi$) subsystem of an electron, that is measured, and the rest of the system that is not measured.

[7] Other important papers pertaining to this issue are: Imry, cond-mat/0202044, Gavish, Levinson & Imry, PRB 62, R10637 (2000); Cohen & Imry, PRB 59, 11143 (1999); Buttiker, cond-mat/0106149 and cond-mat/0105519; Nagaev & Buttiker, cond-mat/0108243; Cedraschi & Buttiker, Annals of Physics (NY) 289, 1 (2001); PRB 63, 165312 (2001); Cedraschi, Ponomarenko & Buttiker, PRL 84, 346 (2000).

– The most important argument against zero temperature decoherence in a system of electrons (fermions) is the apparent role of Pauli exclusion. At T=0, electrons occupy all energy levels up to Fermi energy. Further scattering of individual electrons is prohibited as there is no more scattering states left for an electron to scatter into. Thus a single electron as a part of a system of electrons cannot decohere at T=0 even by the interaction among the particular electron and the rest of electrons. However—as the counter argument goes, the state with interaction cannot be obtained from the non-interacting state of fermions, especially if the interaction is strong. The ground state is no more a pure state in this case; many examples exist to this effect.
– Along the line of the last argument, if electrons decohere at T=0, then they cannot be described as quasi-particles, as in the Fermi liquid theory, which has worked well so far. Thus, they must not decohere at T=0.

The formal contention is based on the calculations that essentially justify or negate the last point. Scattering of a single electron is prohibited as Pauli exclusion prevents the change of its state or wave function. This gives rise to the much debated tanh term in the coth-tanh term for the density of final states for scattering. One group of theorists maintains that cancellation of various contributions to decoherence occurs precisely due to the coth-tanh factor. The other group argues that the cancellation merely reflects the perturbative nature of the calculations and the improper use of the Fermi's golden rule for the calculation of the scattering rate. In a non-perturbative calculation one indeed observes a non-vanishing decoherence rate, consistent with the analysis in the Caldeira-Leggett model. They further argue that the first order expansion in fact agrees with the perturbative calculation.

The Fermi-liquid approach starts from a ground state at T=0, *even in the presence of disorder and electron interaction*. It describes various low-energy properties of metals by low-lying excitations close to this ground state. In this approach, it suffices to treat interaction perturbatively. However, a mixed state at T=0 instead of an idealized many-body pure state is suggested by the observation of temperature independent decoherence. Phenomenologically, the theory of disordered conductors in low dimensions seems to be inadequate in explaining the experimental data. In addition to numerous discrepancies in metallic and insulating systems, high $T_c$ superconductors do not behave as Fermi-liquid metals, making the necessity of addressing this inadequacy even more imminent.

The notion of dephasing in mesoscopic physics needs to be re-examined. Traditionally, one assumes a well-defined phase for the electron wave function, which then acquires small phase shifts ($\ll 2\pi$) due to the coupling to an environment. Averaging over randomness, such as thermal fluctuations, results in the dephasing rate. This prescription is valid for small phase shifts, or when the electron is weakly coupled to the environment, consistent with a perturbative analysis. However, in the strong coupling regime the determination of dephasing rate without the inclusion of the environmental dynamics may be inappropriate as it loses a lot of important physics such as back reaction. Furthermore, it is well known, in the quantum-Brownian-motion models of decoherence that factorization of the initial density matrix into the system (electron) and the environment parts introduces non-unitarity. Considering these conceptual problems, it may be proper to replace the notion of dephasing by that of decoherence, formally described as the decay of off-diagonal terms of the reduced density matrix.

In overall totality, the immediate difficulty that needs to be sufficiently addressed is the unambiguous discrepancy between the perturbative diagrammatic analysis and the density-matrix path-integral approach, based on the quantum-Brownian-motion models.

## 6. ENDNOTE

Quantum decoherence at zero temperature is important to problems in other areas of physics as well. In the foundations of quantum theory, the measurement problem arises because of the necessity of a classical apparatus, and hence a classical theory, to interpret the results of a quantum theory of which it should be a limiting case. The formal theories of decoherence are constructed to explain this problem in addition to the problem of lack of interference or superposition in macroscopic objects. Though most of these models involve the environment in the high temperature limit, zero temperature decoherence within the quantum framework is more essential. The electron decoherence experiments provide an avenue for the verification of these theories.

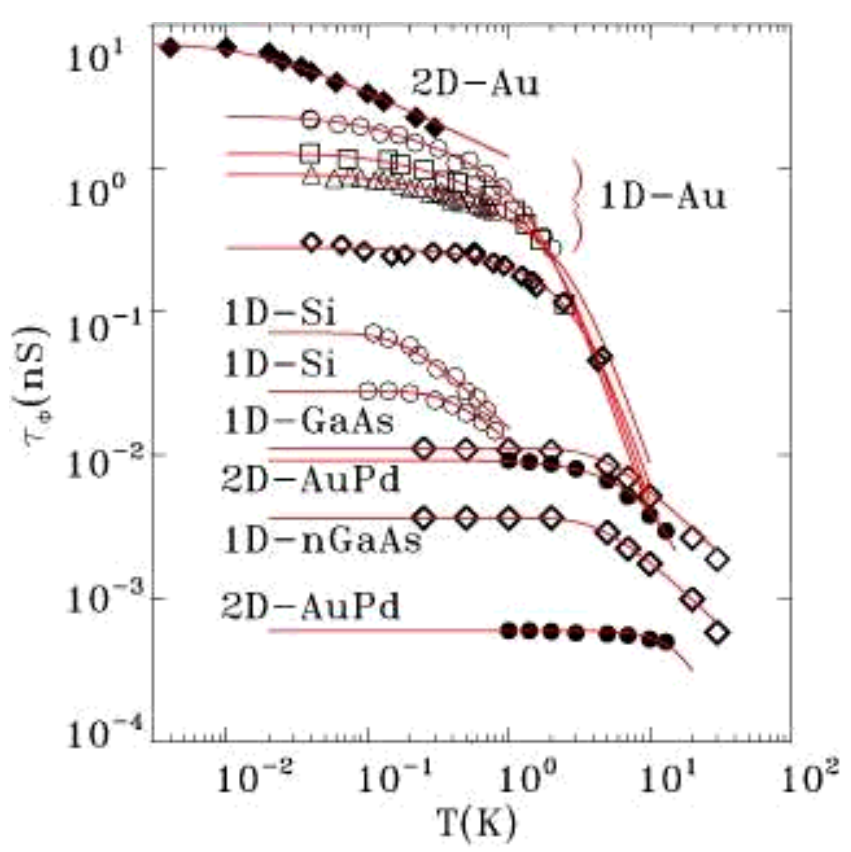

*Figure 19.* A number of 1D and 2D systems showing the saturation of decoherence time.

Quantum information depends crucially on the coherence of the qubits. Like any other quantum system, the qubits also undergo decoherence due to their interaction with the environment. Recent experiments on superconducting qubits find a temperature-independent decoherence rate at low temperatures, consistent with the theoretical predictions based on a Caldeira-Leggett model. Though, the conceptual problem of Pauli exclusion is not relevant to these systems, a limit to coherence of the qubits appears to be fundamental. The debates and controversies on electron decoherence in metals become directly relevant to quantum computation based on fermionic qubits, suggested recently. Recently, it has also been argued that the unification of fundamental interactions may require the zero temperature decoherence of quantum black holes. Many other problems in cosmology and gravity, including particle production in early universe and quantum gravity, also invoke some kind of decoherence mechanism operating at zero temperature.

Interestingly, the reason for which the observed phase coherence saturation effect is extremely important—it goes contrary to the conventional wisdom, at least in condensed matter physics—is the same reason why it is natural to have conceptual difficulties with it. It remains to be seen whether electron-electron interaction (or something else) indeed gives rise to the zero-temperature electron decoherence. Most definitely, more direct measurements of decoherence rate are needed to set the phenomenology on a firm footing. Experiments on various associated phenomena along with their correlation with the saturation are also necessary both for consistency and a better understanding of the anomalies. In the face of current difficulties, the hope is that the experimental and theoretical study of electron decoherence will lead to a better understanding of many unresolved problems in physics.